\documentclass[]{spie}  
\usepackage[]{graphicx}

\title{Calibration and flight performance of the long-slit imaging dual order spectrograph} 

\author{Roxana E. Lupu\supit{a}, Stephan R. McCandliss\supit{a}, Brian Fleming\supit{a}, Kevin France\supit{b}, Paul D. Feldman\supit{a} and Shouleh Nikzad\supit{c}
\skiplinehalf
\supit{a}Department of Physics and Astronomy, Johns Hopkins University, 3400N Charles St., Baltimore, Maryland 21218, USA; \\
\supit{b}Center for Astrophysics and Space Astronomy, University of Colorado, Boulder, Colorado 80309, USA; \\
\supit{c}Advanced Detector and Nano Science Technologies Group, Jet Propulsion Laboratory, 4800 Oak Grove Drive, Pasadena, California 91109, USA
}

\authorinfo{R. E. L. E-mail: roxana@pha.jhu.edu, Telephone: 1 410 516 8485}

\begin{document} 
  \maketitle 

\begin{abstract}
We present a preliminary calibration and flight performance of the Long-Slit Imaging Dual Order Spectrograph (LIDOS), a rocket-borne instrument with a large dynamic range in the 900 - 1700 \AA\ bandpass. The instrument observes UV-bright objects with a CCD channel and fainter nebulosity with an MCP detector. The image quality and the detector quantum efficiencies were determined using the calibration and test equipment at the Johns Hopkins University, and further monitored using an on-board electron-impact calibration lamp. We review results from each of the three flights of the instrument.
 
\end{abstract}



\section{INTRODUCTION}
\label{sec:intro}  

One of the ongoing problems in astrophysics is to constrain the interactions between hot young stars and the nebular material in which they form. Characterization of the properties of gas and dust in such environments using spectroscopy is enhanced by knowledge of the spectral energy distribution of the illuminating star(s) in the far-ultraviolet (far-UV) bandpass (900 -- 1650 \AA). Such knowledge allows us to quantify the relationship between  extinction of the far-UV radiation field by dust and molecular hydrogen (H$_{2}$) fluorescence by simultaneously measuring the spectra of the exciting stars and the scattered light from the surrounding nebular material. The Long-slit Imaging Dual Order Spectrograph (LIDOS) is uniquely equipped for this task, employing a charge-coupled-device (CCD) channel to observe UV-bright objects and an microchannel plate (MCP) detector for the fainter emission. The long-slit configuration combined with the resolving power of the telescope, allows for the extraction of information as a function of the angular offset from the point source. The instrument is designed to be flown on a sounding rocket in order to avoid UV attenuation by the atmosphere. LIDOS has flown three times and has obtained data on the $\gamma$~Cas/IC~63 system (36.208UG, December 2003), the Trifid Nebula (36.220UG, August 2007) and the Orion Nebula (36.243UG, January 2008). The targets were chosen to provide a proving ground for the instrument, with stars that would have exceeded the brightness limits of the $Far-Ultraviolet$ $Spectroscopic$ $Explorer$ ($FUSE$). All flights were successful, with the MCP collecting data on all three objects, and the CCD acquiring the spectrum of $\gamma$~Cas\cite{France:05} and $\theta^1$~Ori~C, the brightest star in the central region of the Orion nebula. All flights were preceded and followed by refurbishing and end-to-end characterization of the instrument in the facilities developed at the Johns Hopkins University. The status of the instrument was monitored during all phases of the integration through the use of an on-board e$^{-}$-impact lamp\cite{McCandliss:03} mounted on the side of the spectrograph and a similar lamp on the shutter door of the telescope.

The scope of this paper is to review the characteristics of the LIDOS instrument as measured during flight and during extensive laboratory calibrations, and to show the preliminary results obtained from the flight data. It serves as a performance study of the LIDOS design and calibration techniques for future UV instruments, and as a reference for the data reduction. A brief description of the instrument is given in Section 2, followed by the description of the calibration procedures and the derived instrument characteristics in Section 3. The flight performance is presented in Section 4 and the flight results follow in Section 5.

\section{LIDOS and Payload Description}

The LIDOS payload consists of three sections: the telescope, the spectrograph and the electronics. The 40 cm diameter f/16 Dall-Kirkham telescope is composed of a three-point mount for the primary mirror and an Invar tube holding the secondary and the star-tracker mount. The shutter door has been modified to accommodate an electron-impact calibration lamp\cite{McCandliss:03,France:02} used to monitor the degradation of the optics. The telescope is focused separately prior to the attachment of the spectrograph. The entire telescope-spectrograph chamber is evacuated to p~$\le$~10$^{-5}$~torr prior to flight by an external pump stack.

\begin{figure}
   \begin{center}
   \begin{tabular}{c}
   \includegraphics[angle=90,height=3.5cm,clip]{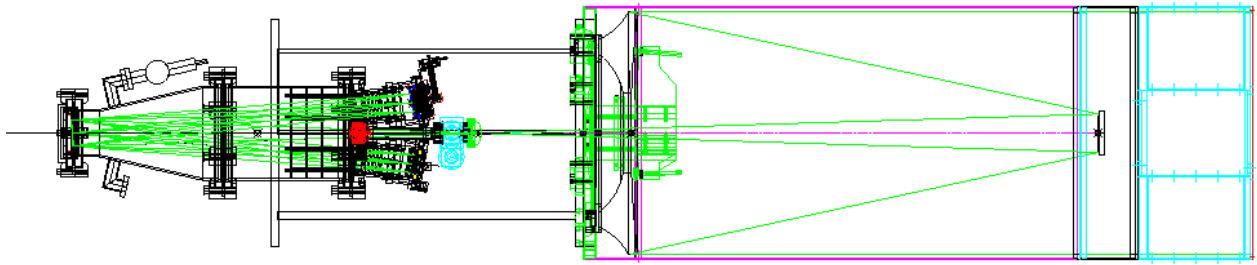}
   \end{tabular}
   \end{center}
   \caption[example] 
   { \label{lidos} 
Optical layout of the LIDOS and telescope assembly. The telescope shutter door is at the right margin of the drawing. The on-board calibration lamp and Xybion camera, mounted on the spectrograph section at the left, are not shown.}
   \end{figure}

The light entering the telescope comes to a focus at the 9.25\arcsec$\times$330\arcsec\ slit of our off-Rowland circle spectrograph design. The detectors are situated inward accordingly, to compensate for the increase in path length. The slit is cut into a tilted flat mirror (slitjaw) that reflects the acquired stellar field (20\arcmin\ field-of-view) into a Xybion TV camera, enabling real-time maneuvers during flight. The Xybion camera is an image intensified CCD and is subject to saturation on bright sources, depending on gain settings. The details of the spectrograph design and construction have been given in Ref.~\citenum{McCandliss:03}. The optical layout of the telescope and spectrograph assembly is shown in Figure \ref{lidos}.

The positive and negative first order spectra generated by a holographically ruled CVD SiC diffraction grating are recorded by two primary science detectors: a microchannel plate stack with a CsI photocathode and a double delay-line anode\cite{siegmund:93}, and a CCD with a delta-doped backside\cite{Nikzad:00,Nikzad:94}. The CCD can accommodate higher count rates, limited by the full-well limit and the shortest exposure time. The CCD dark current at room temperature overfills the A/D converter in a bias frame. In order to reach its lowest brightness limits, the CCD has to operate at 200~K. This is achieved by a Joule-Thompson cryostat working in conjunction with a thermo-electric cooler. At the working temperature the detection threshold limit is set by the read noise. The MCP detector on the other hand has a low background equivalent flux (BEF) which allows for the detection of fluxes three orders of magnitude lower than the CCD. At high count rates, however, the MCP is limited by dead time corrections due to the electronics speed and charge depletion in the pores, the response becoming highly non-linear. Part of the MCP was covered with a 1~mm wide occulting strip to allow simultaneous observation of a bright source with the CCD and of the surrounding material with the MCP. 

The spectrograph and detector enclosure is sealed by a gate valve and maintained under 10$^{-8}$~torr by a non-evaporable getter pump and a vacuum-ion pump to prevent contamination. The gate valve is activated during flight by a dedicated motor. A $\pm$~100~V repeller grid located just in front of the gate-valve is used to reject charged particles from the spectrograph. A windowless electron-impact X-ray lamp has been permanently added to the spectrograph housing and, by illuminating the grating via a horseshoe mirror, allows for checking the MCP response on the fly. The lamp consists of a tungsten filament that acts as a source of free electrons under an applied current, and a thick tungsten target maintained at a bias of $\approx$~1000 V.

The electronics section includes some of the detector amplifiers and A/D converters, the motor controllers, power supplies, and the on-board telemetry (TM) interface. The telemetry interface collects the raw pulses from the MCP detector and downlinks the data from the CCD channel during flight. Housekeeping data and uplink commands for real-time exercise of the fine pointing, detector high voltage, spectrometer ion-pump, lamp power and CCD timing and  exposure commands are also managed by TM using a 10 Mbit~s$^{-1}$ link.

The pointing of the telescope is achieved via a ST-5000 star tracker mounted aft of the secondary mirror. The star tracker signals are used in conjunction with a LN-251 gyro by the Attitude Control System (ACS) for the target acquisition and pointing stability. Fine guiding maneuvers are achieved via the Command Uplink System (CUS) using the on-board Xybion camera. The ACS for the 36.208 flight was an Aerojet Mark VI which was replaced by the Celestial ACS developed by the NASA Sounding Rocket Operations Contract (NSROC) for the subsequent flights (36.220 and 36.243). These flights have provided testing opportunities for the ST-5000 and the Celestial ACS, and we review our perspective on their performance in \S~4.

    \begin{figure}
   \begin{center}
   \begin{minipage}[t]{7.5cm}
   \begin{center}
     \includegraphics[angle=90,height=5cm,clip]{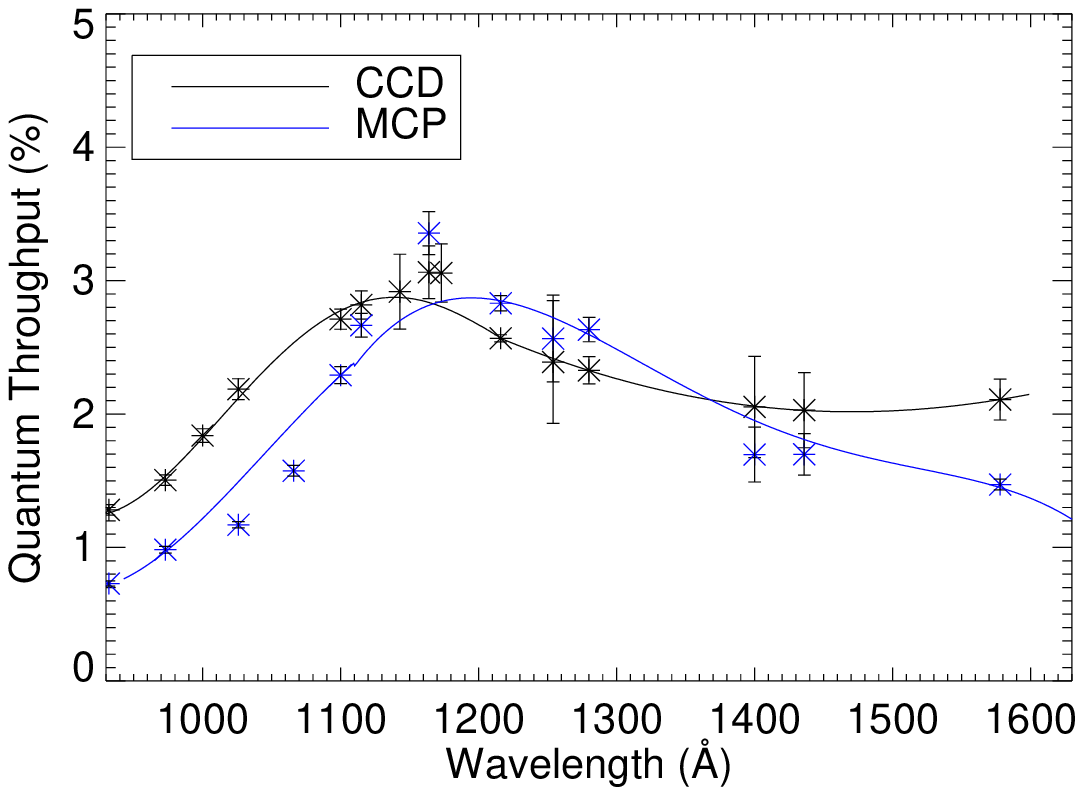}
      \end{center}
      \caption[example]   { \label{qe} 
Quantum throughput of the LIDOS instrument for each channel. The symbols represent direct measurements and the continuous lines are polynomial fits to the data. }
 \end{minipage}
 \hspace{1cm}
 \begin{minipage}[t]{7.5cm}
   \begin{center}
 \includegraphics[angle=90,height=5cm,clip]{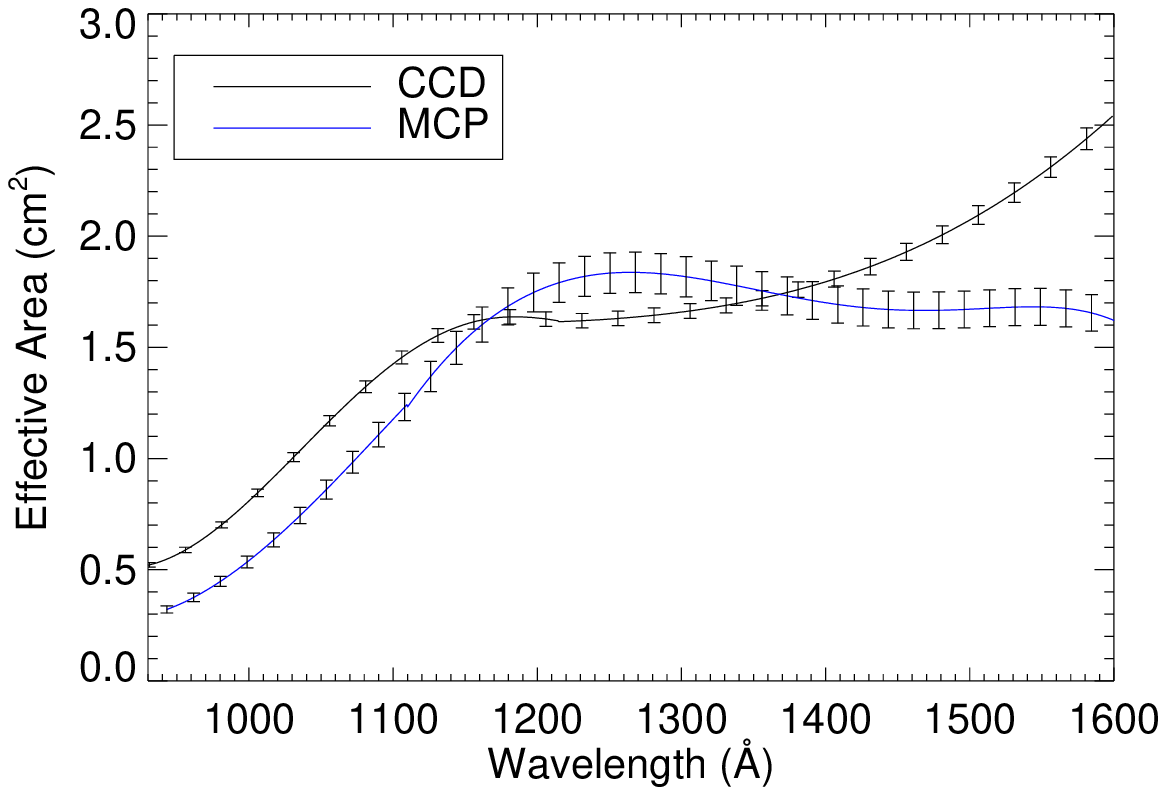}
 \end{center}
  \caption[example]  { \label{effarea} 
Total effective area of the telescope and spectrograph assembly, estimated with the in-flight value for the mirror reflectivity.}
\end{minipage}
\end{center}
    \end{figure}
    
    \begin{figure}[b]
   \begin{center}
   \begin{tabular}{c}
   \includegraphics[width=14cm,clip]{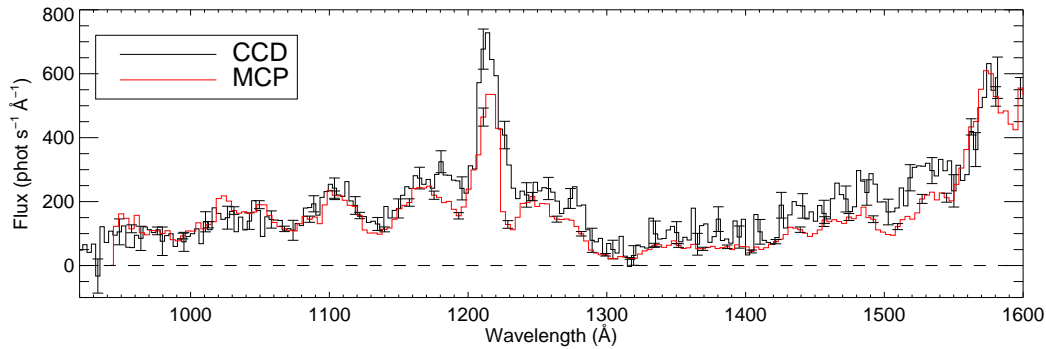}
   \end{tabular}
   \end{center}
   \caption[example] 
   { \label{xcal} 
Cross-calibration spectrum using a hydrogen discharge lamp.}
   \end{figure}

\section{Calibration and Testing}

    \begin{figure}[t]
   \begin{center}
   \begin{tabular}{c}
   \includegraphics[height=5cm,clip]{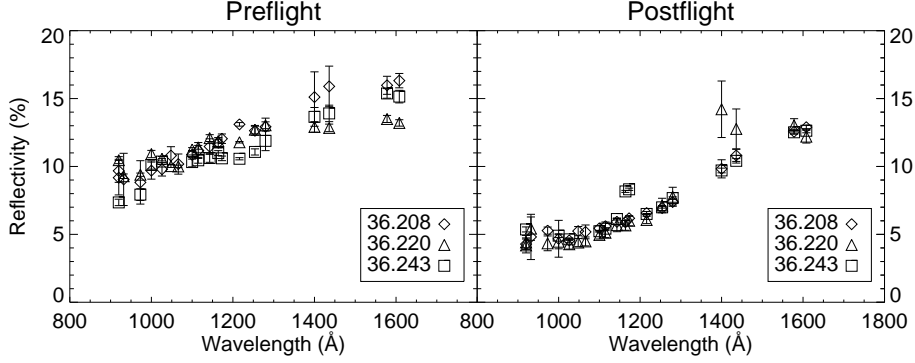}
   \end{tabular}
   \end{center}
   \caption[example] 
   { \label{refl} 
Pre- and post-flight measurements of the telescope total reflectivity.}

   \end{figure} 
\begin{figure}[hb]
   \begin{center}
   \begin{minipage}[t]{7.cm}
   \begin{center}
     \includegraphics[angle=90,width=7cm,clip]{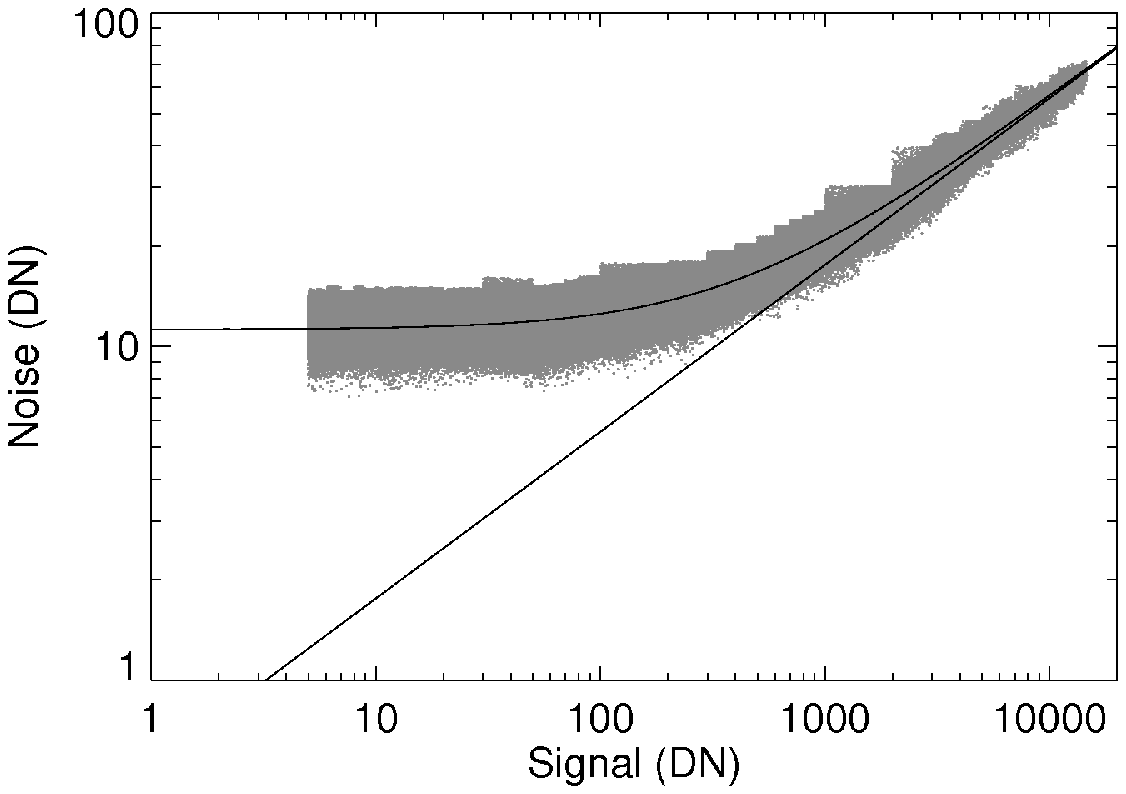}
      \end{center}
      \caption[example]   { \label{ptc} 
CCD photon transfer curve previous to the 36.243UG flight. The fit for 3.2~e$^{-}$/DN and 11.3~DN read noise is shown in black, and the data points are shown in gray.}
 \end{minipage}
 \hspace{1cm}
 \begin{minipage}[t]{7.cm}
   \begin{center}
 \includegraphics[angle=90,width=7cm,clip]{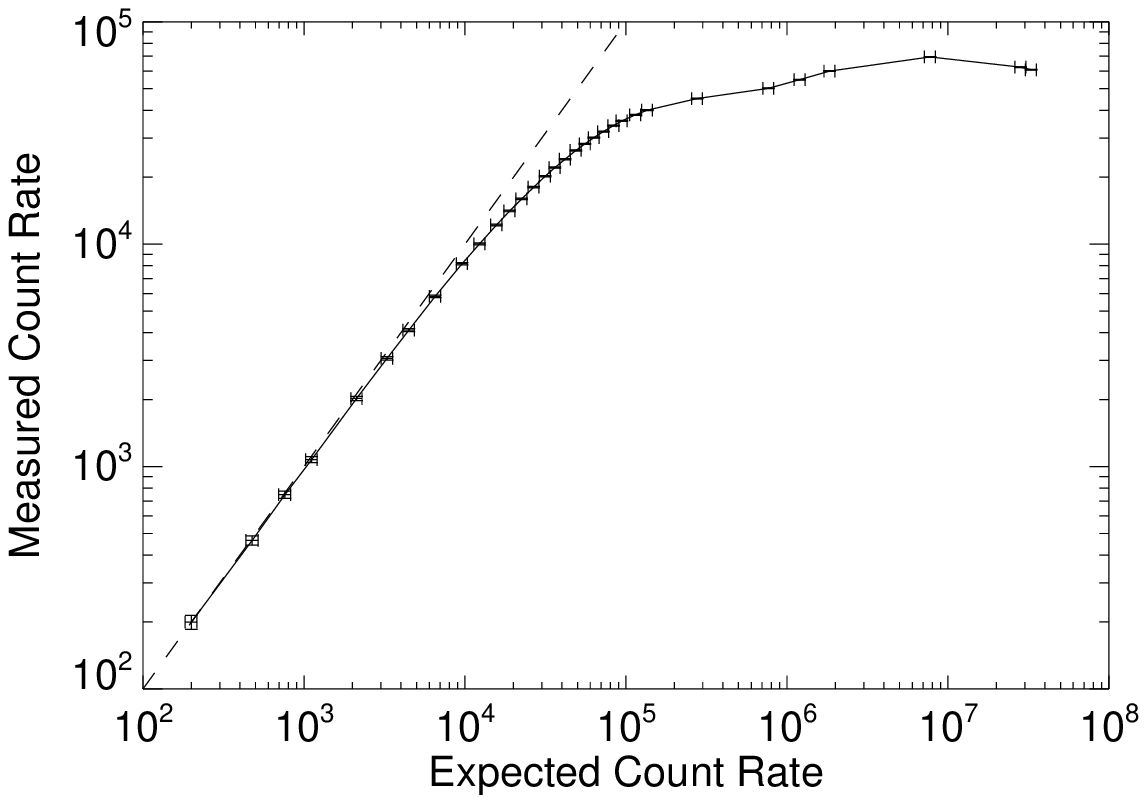}
 \end{center}
  \caption[example]  { \label{linear} 
Measurement of the MCP count rate linearity after the 36.243UG flight. A linear dependence is shown with a dashed line.}
\end{minipage}
\end{center}
    \end{figure}

\paragraph{Quantum Throughput.} The quantum throughput of the two channels was determined and cross-verified using the Calibration and Test Equipment (CTE) at the Johns Hopkins University. The facility comprises a vacuum far-UV monochromator fed by a windowless hollow-cathode gas discharge lamp, separated by an interface valve from a large vacuum chamber housing the instrument under study. Wavelength dependent characterizations are made using the strong lines of molecular hydrogen and argon discharge. The CTE setup also allows us to test and adjust the best values for the CCD gain and operating temperature, and the MCP high voltage.  

The quantum throughput of LIDOS is a product of the quantum efficiencies of the detectors, the grating groove efficiency, and the transmission of the repeller grid situated at the entrance of the spectrograph. The total quantum throughput is derived by using the total flux measured from a pair of UV detectors, a microchannel plate working between 920 and 1216~\AA, and a photomultiplier tube Model~542 with a CsI photocathode at wavelengths longer than 1216~\AA. These secondary standards are calibrated beforehand against a set of laboratory primary standards. these standards are a pair of NIST calibrated ultraviolet photodiodes: an Al$_{2}$O$_{3}$ detector for the 500--1216~\AA\ bandpass (SN~239), and a windowed Cs$_{2}$Te detector working longward of 1160~\AA\ (SN~682-3). The data from the LIDOS detectors is recorded for five choices of the beam incidence angle relative to the grating. The MCP measurement is performed for two different positions on the detector, one on each side the occulting strip. There was no observed variation of the spectrograph response with incidence angle, only a few percent difference between the two sides of the MCP detector, possibly due to beam obscuration by the strip or the repeller grids. The average throughputs of the two channels are shown in Figure~\ref{qe}. The MCP calibration data is supplemented with the spectrum from the electron impact X-ray lamp to extend the calibration points across the bandpass using the shape of the continuum bremsstrahlung emission\cite{Ajello:88}. Using the zero order of the monochromator to generate an illumination level that gives a linear measurable response in both detectors, we investigated the possibility of cross-calibrating the two channels. The comparison between the two spectra is shown in Figure~\ref{xcal} and reflects the calibration mismatch between the two channels. The whole calibration procedure was repeated at the end of the flight to account for a possible variation of the instrument characteristics, but a significant change was not observed. 
\vspace{-0.2in}
\paragraph{Effective Area.} The total reflectivity of the telescope mirrors in the far-UV was measured as a function of wavelength pre-flight and post-flight using the same light source in the CTE. The reflectivity was measured again after the flight to monitor degradation. The final result is an average from three different incident positions of the monochromator beam on the primary mirror. As shown in Figure~\ref{refl}, the reflectivity drops by about a factor of two between measurements at the shortest wavelengths. The total effective area for the two LIDOS channels was derived as a product of the total efficiency, mirror reflectivity, and clear area of the telescope. In Figure~\ref{effarea} we show the derived effective area using the degraded reflectivity which should be an accurate measure of the instrument performance during flight. The effective area for the CCD channel uses the conversion factor E/3.65 for the electron yield per incoming photon, where E is the energy of the photon in eV\cite{Janesick:87}. The number of electrons corresponding to one CCD digital number (DN) was measured from the photon transfer curve, as described below.
\vspace{-0.2in}
\paragraph{CCD Photon Transfer.} We developed a procedure for using the CCD calibration images to construct a photon transfer curve\cite{Janesick:87,Morrissey:95} and infer the detector read noise and the number of electrons generated per DN as a function of wavelength. Two images are taken at each wavelength at the same illumination. The images of the averaged intensities and the difference images are divided into boxes covering uniformly illuminated regions over which the statistics are collected. The boxes cover each zone of illumination, from the background level to the peak of the spot generated by the monochromator. The scatter in the correlation between the signal and the noise will be dependent on the box size, but we found it is possible to obtain a good fit and low scatter with boxes containing a sufficient number of pixels for reliable statistics. Areas with large standard deviations compared to other areas with a similar signal level are discarded to avoid large gradients of the signal in the non-uniformly illuminated images. The photon transfer curve derived with the last settings for gain and operating temperature before the 36.243UG flight is shown in Figure~\ref{ptc}. The curve represents a fit using all the points across the wavelength range. There was no observed wavelength dependence of the number of electrons per DN. The best fit results in a read-noise of 11.288$\pm$0.002~DN and a conversion factor of 3.2184$\pm$0.0006~e$^{-}$/DN. 
\vspace{-0.2in}
\paragraph{MCP linearity.} After the photon detection efficiency measurements in the CTE we performed a linearity test to determine the dead time correction for the MCP count rate. Using the zero order of the monochromator we were able to obtain count rates as high as 6.5$\times$10$^{4}$ counts s$^{-1}$. In Figure~\ref{linear} we show the MCP linearity curve obtained from the internal detector counter bypassing the data acquisition programs that cannot keep up with such a large count rate. By comparing the curve with a linear dependence, as indicated by the dashed line in the figure, we estimate that the detector starts to become non-linear around 1-2$\times$10$^{3}$~counts~s$^{-1}$. This procedure, however, does not take into account the spectral energy distribution of the lamp and it is likely to underestimate the magnitude of the count rate correction. The hydrogen spectrum exhibits a very strong Ly$\alpha$ feature which will likely saturate a small area on the detector much faster than the rest, while the photomultiplier tube used as the linear comparison detector measures the integrated light over its entire surface. Therefore, the MCP count rate will be determined by a combination of the dead time of the electronics and charge depletion on the detector. This process leads to the saturation of the measured MCP count rate, while masking the turnover due to the counting speed of the electronics at the same time. The lower count rate due to charge depletion would allow the electronics to keep the pace with a higher measured incoming flux, thus leading to an underestimation of the count rate correction for the flight data, when charge depletion was not a factor due to the flat nature of the spectrum.
\vspace{-0.2in}
\paragraph{Wavelength calibration.} Wavelength calibration was performed using a spectrum taken in a similar setup, with stronger lines obtained using a calibrated air leak in the vacuum chamber. The main features due to dissociation products of nitrogen, water vapors, and carbon dioxide (N$_{2}$, \ion{N}{1}, \ion{H}{1}, \ion{O}{1}, CO, \ion{C}{1}, \ion{C}{2}), were identified and their positions determined by gaussian centroids. The wavelength scale was established from an iterative polynomial fit to the derived points.
\vspace{-0.2in}
\paragraph{Focus.} Upon assembly, the telescope focus was adjusted and measured to $\pm$1~mm using a knife-edge setup, as described in Ref.~\citenum{Burgh:01}, and an end-to-end testing was performed using the electron-impact soft X-ray lamp mounted on the shutter door at the aft end of the telescope. This measurement served as a baseline for monitoring the payload contamination during various phases of the mission. 
\vspace{-0.2in}

 \begin{figure}
   \begin{center}
   \begin{tabular}{c}
   \includegraphics[height=6cm,clip]{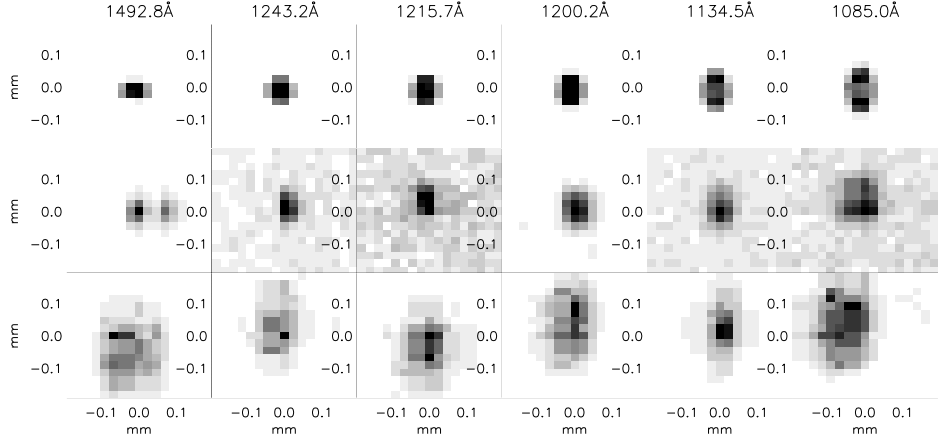}
   \end{tabular}
   \end{center}
   \caption[example] 
   { \label{psf} 
PSF of the LIDOS and telescope assembly as a function of wavelength. The top row shows the predicted PSF (see text), the center row contains the CCD images, and the bottom row shows the MCP data.}
   \end{figure}

\begin{figure}[p]
   \begin{center}
   
   \includegraphics[width=17cm,clip]{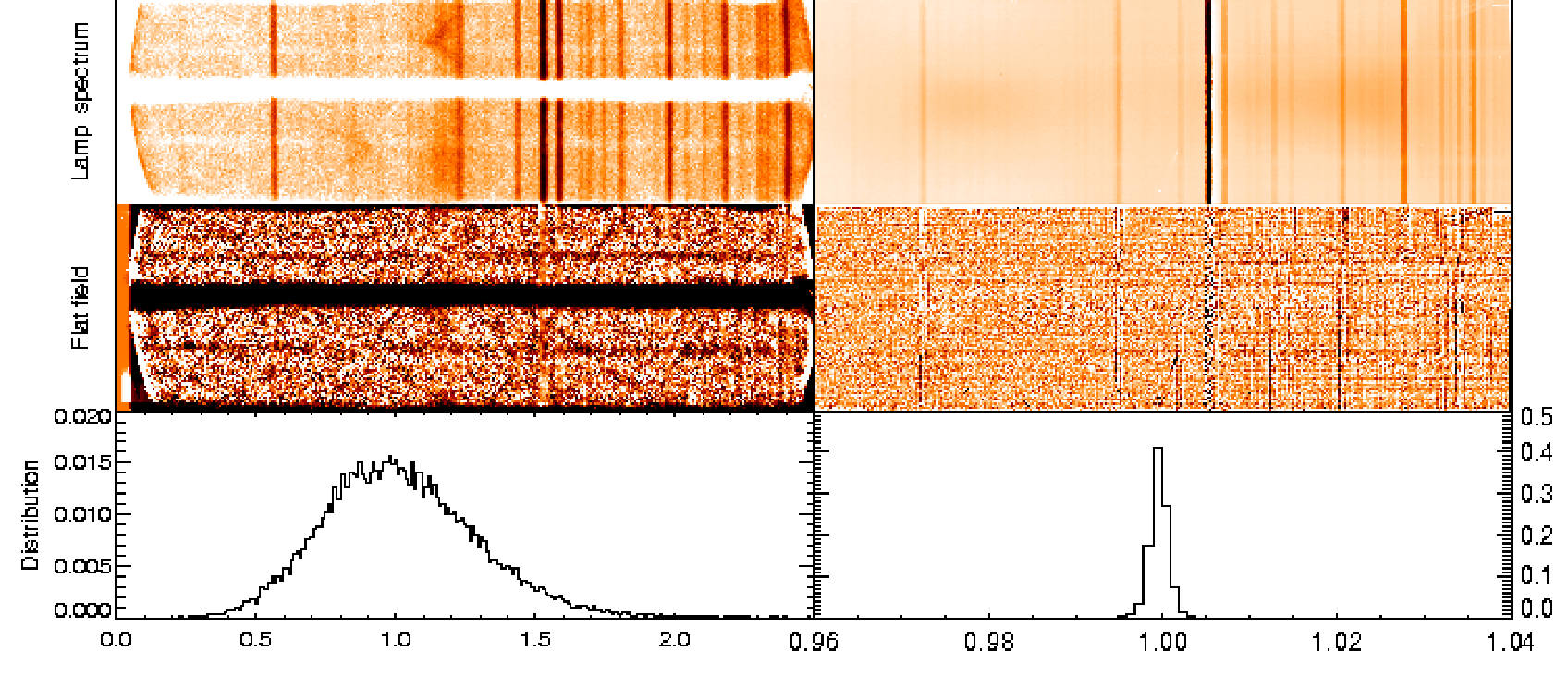}
   
   \end{center}
   \caption[example] 
   { \label{flats} 
Spectrum of the electron-impact lamp mounted in front of the entrance aperture of the spectrograph and post-processing flat fields for the MCP (left) and CCD (right). Histograms of the flat field variation are included at the bottom.}
   \end{figure}

\paragraph{Point Spread Function.} The final end-to-end point spread function (PSF) of the instrument was measured as a function of wavelength using a coronal gas discharge lamp mounted on a vacuum collimator. The vacuum collimator, described previously in Ref.~\citenum{Burgh:01}, is fed by a 20~$\mu$m pinhole situated on a motorized stage with three axis maneuverability. The size of the pinhole is chosen to generate a spot at the expected resolution limit of the optics. The stage motion allows for the spectrograph slit mapping and focus verification in UV light. The position of the focus is measured by taking a series of exposures while moving the pinhole stage along the optical axis. The angles of the mirrors are adjusted by determining and measuring the magnitude of the coma in the recorded images. The procedure is repeated until the optical quality is satisfactory. A comparison between the expected PSF of the instrument and the best observed PSF with the two detectors is shown in Figure~\ref{psf}. The predicted PSF is generated by ray-tracing the full testing setup, including the vacuum collimator, the telescope and the spectrograph, using the 20~$\mu$m pinhole as the light source and defocusing the telescope by the same amount as measured (vertex-to-vertex change of 0.02~mm). All images have been rebinned to a 24~$\mu$m pixel size, equal to the CCD scale. The measured spatial resolution of the MCP using the line spread function from the occulting strip is 6.05$\pm$0.20 pixels, or 6.94$\pm$0.23\arcsec\ at Ly~$\alpha$. The best measured CCD spatial resolution for the 20~$\mu$m pinhole was 1.52$\pm$0.05 pixels or 1.41$\pm$0.05\arcsec\ at Ly~$\alpha$. The spatial resolution in arcseconds varies slightly across the wavelength range due to the demagnification factor introduced by the off-Rowland circle design. Moreover, even at best focus, the spatial resolution varies with wavelength due to the growth of the astigmatism away from the wavelength at which the grating is corrected (1275~\AA). We estimated the spectral resolution for a filled aperture and a point source from the deep flat field spectra and the slit line spread function. The MCP spectral resolution varied from 7.38$\pm$0.10 pixels for the filled aperture to 3.56$\pm$0.06 pixels for point sources (5.27$\pm$0.13~\AA\ and 2.54$\pm$0.07~\AA, respectively), while the CCD spectral resolution was measured to be 7.81$\pm$0.36 pixels for the filled slit and 2.31$\pm$0.04 pixels for point sources (4.83$\pm$0.25~\AA\ and 1.43$\pm$0.04~\AA, respectively).
\vspace{-0.2in}
\paragraph{Flat-fields.} We used an electron-impact windowless X-ray lamp placed in front of the entrance slit of the spectrograph to construct the flat field response of the two detectors. In Figure~\ref{flats} we show the deep exposures taken with the two detectors and the derived flat field images. The MCP image consists of eight exposures totaling 6397~s of data acquisition, while the CCD spectrum is composed of 68 exposures, 60~s each, for a total of 4080~s. Each raw image was dark subtracted and divided by the median value in the spatial direction to remove the spectral features. The large scale features due to non-uniformities in the illumination, or scattered white light in the case of the CCD, are removed by local median division using the largest box size tolerated by gradients across the image. The resulting variation in the intensity due to flat field corrections amounts to less than 5\% for the CCD and as high as 50\% for the MCP due to the grid lines. A source of uncertainty in the CCD flat fielding is the vignetting profile. The shadow cast by the repeller grids at the entrance of the spectrograph shows variation with signal strength and requires corrections on a case-by-case basis.

\section{Flight Performance}

\begin{figure}[p]
   \begin{center}
   \begin{tabular}{cc}
   \includegraphics[width=11.cm,clip]{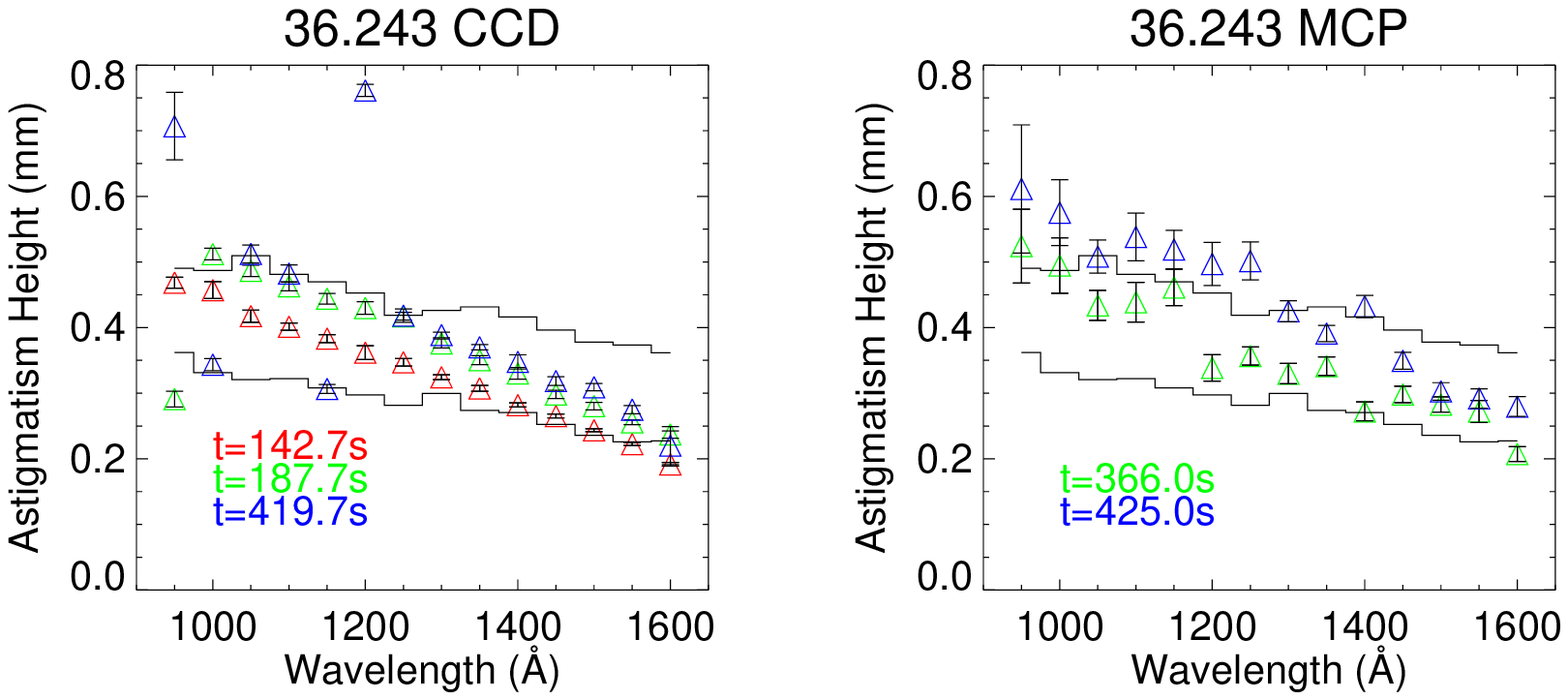}
   \includegraphics[width=5.5cm,clip]{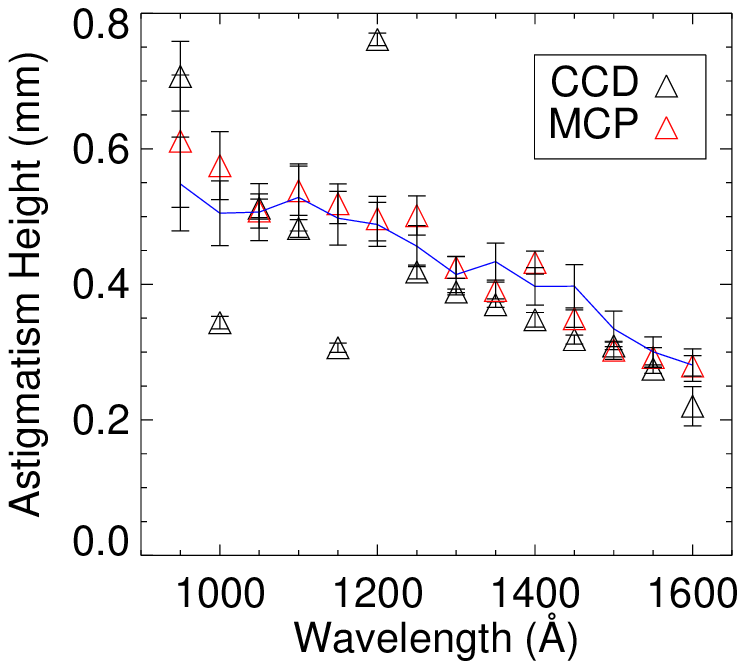}
   \end{tabular}
   \end{center}
   \caption[example] 
   { \label{rlpsf} 
Measured astigmatism height for the CCD (left) and MCP (center) flight data compared to raytrace predictions for an out-of-focus telescope (vertex-to-vertex change of 0.13~mm and 0.19~mm for the lower and upper black lines, respectively), and to post-flight measurements using the vacuum collimator (right, laboratory data in blue). }
   \end{figure}

  \begin{figure}[p]
   \begin{center}
   \begin{tabular}{c}
   \includegraphics[width=16.5cm,clip]{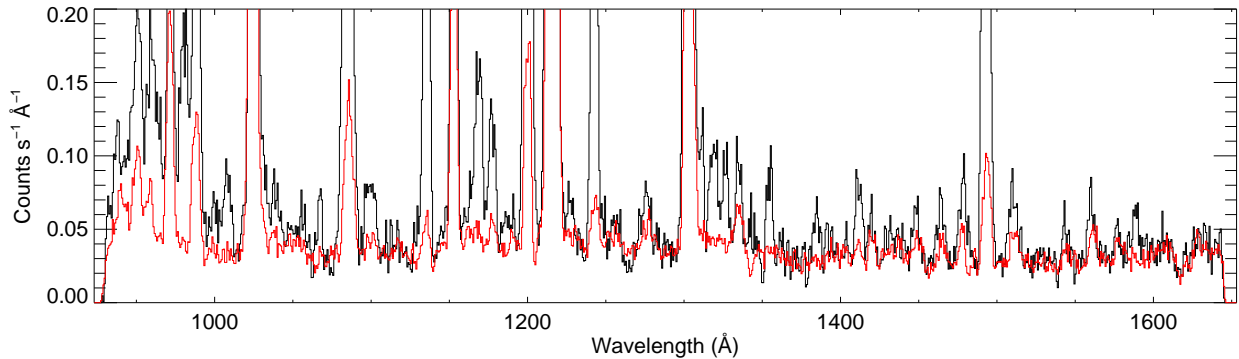}
   \end{tabular}
   \end{center}
   \caption[example] 
   { \label{xlamp} 
Pre- and post-flight MCP data (black and red, respectively) taken using an electron-impact soft X-ray lamp mounted on the telescope shutter door.}
   \end{figure}

LIDOS has almost 400~s of observing time during flight. The experiment gate valve opens 80~s after take-off, followed by the turn on of the Xybion camera. The MCP high voltage turns on at T+102~s, before arriving at the first target at T+112~s. Fine real-time maneuvers are performed and data is recorded by both detectors. The on-board calibration lamp is turned on at the end of each flight, after closing the spectrograph gate valve at T+504~s, to verify the nominal behavior of the MCP. The CCD is used at the same time to take bias frames until the power turn-off at T+552~s. The main sources of concern for the flight performance were related to the CCD cooling system and the stability of the telescope focus. Freeze-ups of the gas flow through the Joule-Thompson cryostat were problematic for maintaining the detector temperature. The thermal background noise during the 36.220UG flight was too high to allow a detection of the stellar spectrum. Extensive testing and a better understanding of the cooling behavior helped the successful CCD observations during the 36.243UG flight, when multiple stellar spectra were recorded. The cooling issues put the largest constraints on the go/no-go launch decisions. 

The focus of the telescope was measured and adjusted pre-flight in both air and  vacuum. After both 36.220UG and 36.243UG flights, however, the analysis of point-source spectra revealed a height of the spectrum in the spatial direction higher than that measured for a simulated point source fed through the vacuum collimator. The intrinsic height of the spectral image is a consequence of the astigmatism, and is a function of wavelength. In the LIDOS instrument the astigmatism is corrected at a single wavelength (1275~\AA\ at nominal focus) by the toroidal figure of the grating, ensuring an equal correction both $\pm$1 orders. The residual astigmatism increases in each direction away from the corrected wavelength. As the position of the focal plane of the telescope changes, the detector will probe a different location relative to the sagittal and tangential foci, causing an apparent shift of the minimum spot height to a different wavelength, and the image size will grow accordingly.

We used the measured height of the point source spectrum as a function of wavelength to constrain the amount of change in the vertex-to-vertex distance of the telescope. We compared the measured astigmatism height as a function of wavelength in the flight CCD and MCP images to the predicted astigmatism height obtained by ray-tracing the telescope-spectrograph assembly with a point source at infinity and a set of values for the vertex-to-vertex distance change ($\Delta_{vv}$). The first two panels of Figure~\ref{rlpsf} show the measured versus simulated wavelength dependence of the astigmatism height for the CCD and the MCP, respectively. The different colors designate different time intervals used for data extraction. The start time for each interval is indicated in each figure. The ray-trace simulations are represented by the continuous black lies, with the lower one corresponding to $\Delta_{vv}$=0.13~mm and the upper one to $\Delta_{vv}$=0.19~mm. The difference in slope is due to the position of the star relative to the optical axis. Based on these measurements we find a slight variation of $\Delta_{vv}$ during flight, and estimate the average value of $\Delta_{vv}$ to 0.16~mm. Direct post-flight measurements of the telescope focus using both the vacuum collimator and an auto-collimation procedure reported a $\Delta_{vv}$ of only 0.035~mm. We were able to reproduce an astigmatism height similar to the one observed in flight only by moving the light source 2.5~mm away from the focus of the vacuum collimator. This comparison is shown in the last panel of Figure~\ref{rlpsf}. The light source displacement is equivalent to an increase in $\Delta_{vv}$ by 0.1~mm, for a total $\Delta_{vv}$ of 0.135~mm, confirming the focus change during flight. A similar result has been established for the 36.220UG flight.

   \begin{figure}[p]
   \begin{center}
   \begin{minipage}[t]{7cm}
   \begin{center}
     \includegraphics[angle=0,height=6cm,clip]{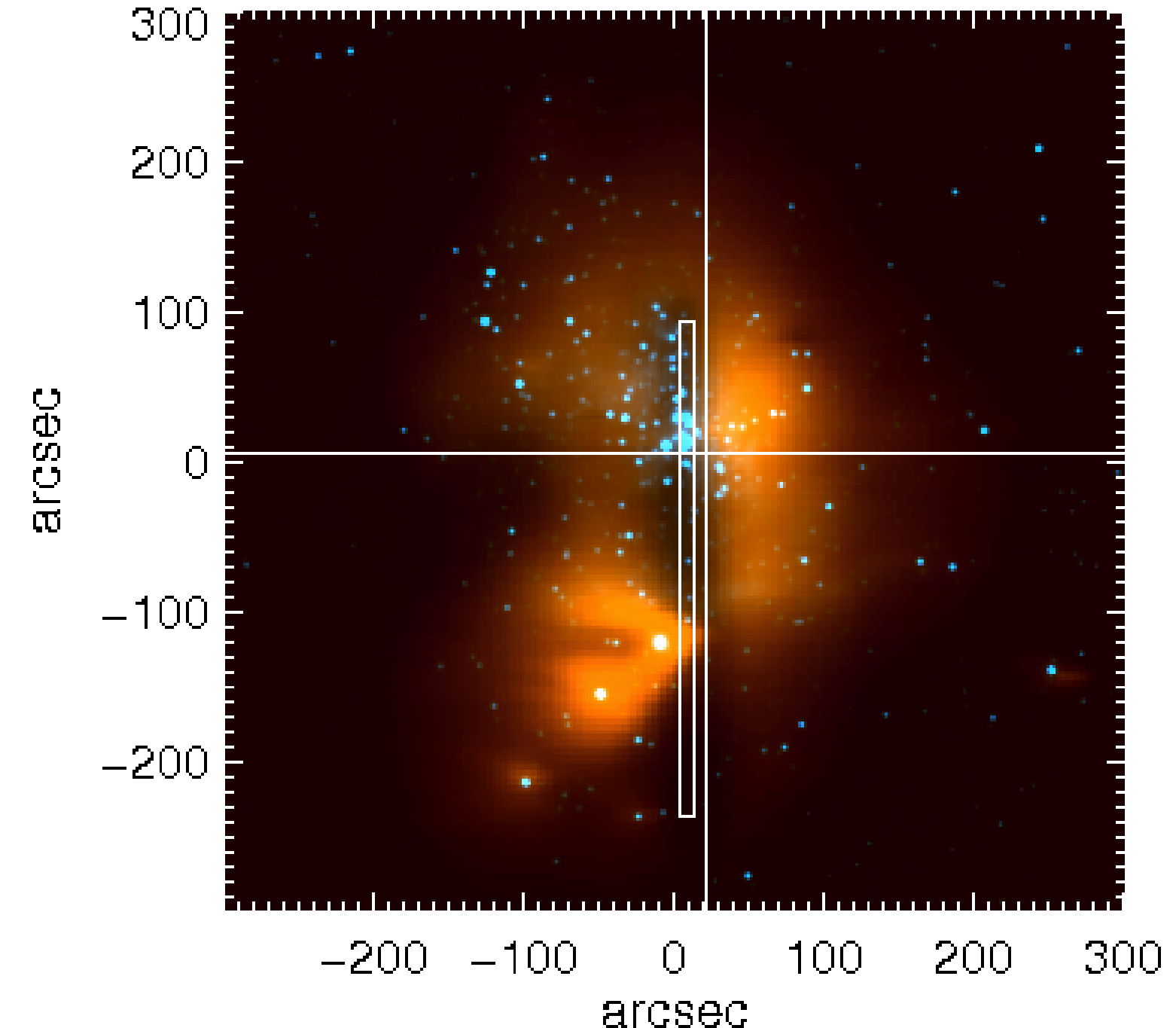}
      \end{center}
      \caption[example]   { \label{xybion} 
Image of the Orion Trapezium region from the on-board Xybion camera (red) overlaid on a 2MASS H-band image of the same field (blue). The ACS pointing is indicated by the crosshairs.}
 \end{minipage}
 \hspace{1cm}
 \begin{minipage}[t]{7cm}
   \begin{center}
 \includegraphics[angle=0,height=6cm,clip]{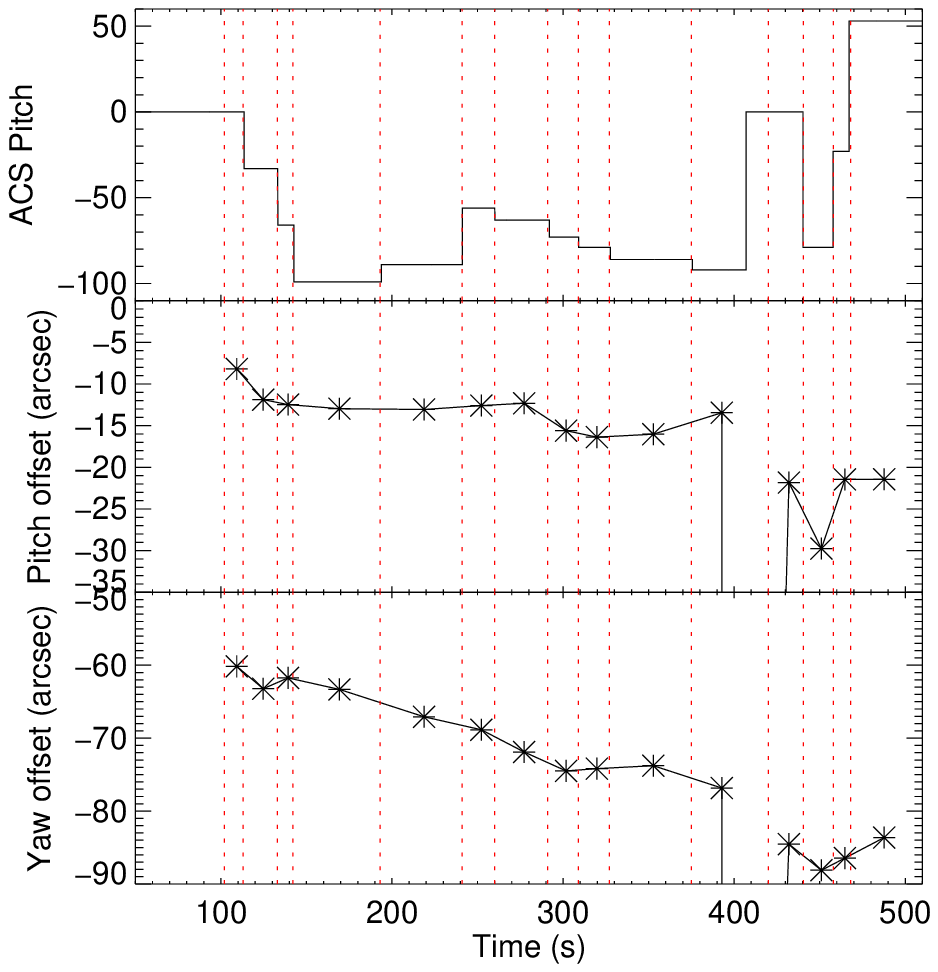}
 \end{center}
  \caption[example]  { \label{acsoff} 
The offset between the ST-5000 star-tracker and the science instrument as a function of time during flight. The intervals correspond to different stable pointing positions, as indicated by the ACS pitch value.}
\end{minipage}
\end{center}
    \end{figure}
       \begin{figure}[p]
   \begin{center}
   \begin{tabular}{c}
   \includegraphics[angle=90,width=16.5cm,clip]{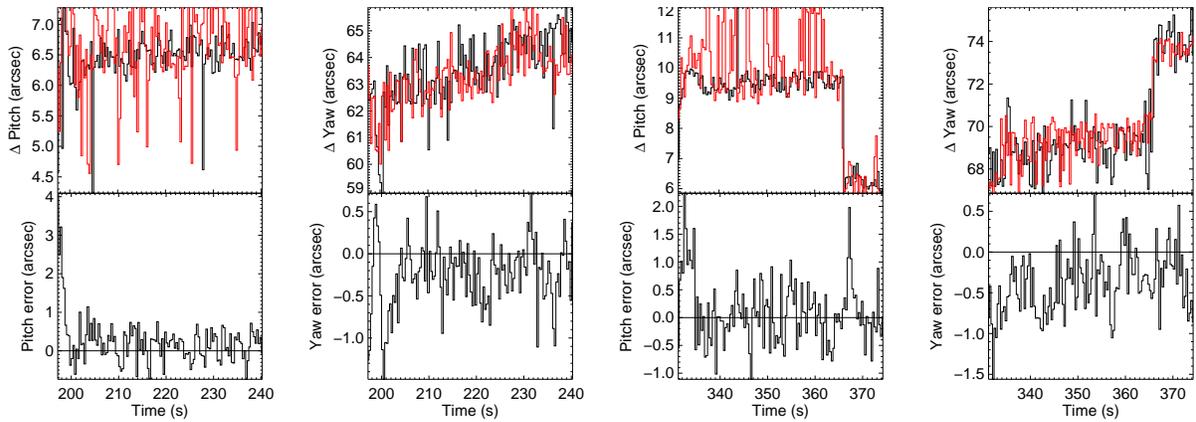}
   \end{tabular}
   \end{center}
   \caption[example] 
   { \label{pshift} 
Top: differences between the coordinates given by ACS and those inferred from the centroids of two stars in the Xybion images (red and black), for two different time intervals (see text). Bottom: pointing stability monitor in the ACS data. }
   \end{figure} 
 \begin{figure}[p]
   \begin{center}
   \begin{tabular}{cc}
   \includegraphics[width=11cm,clip]{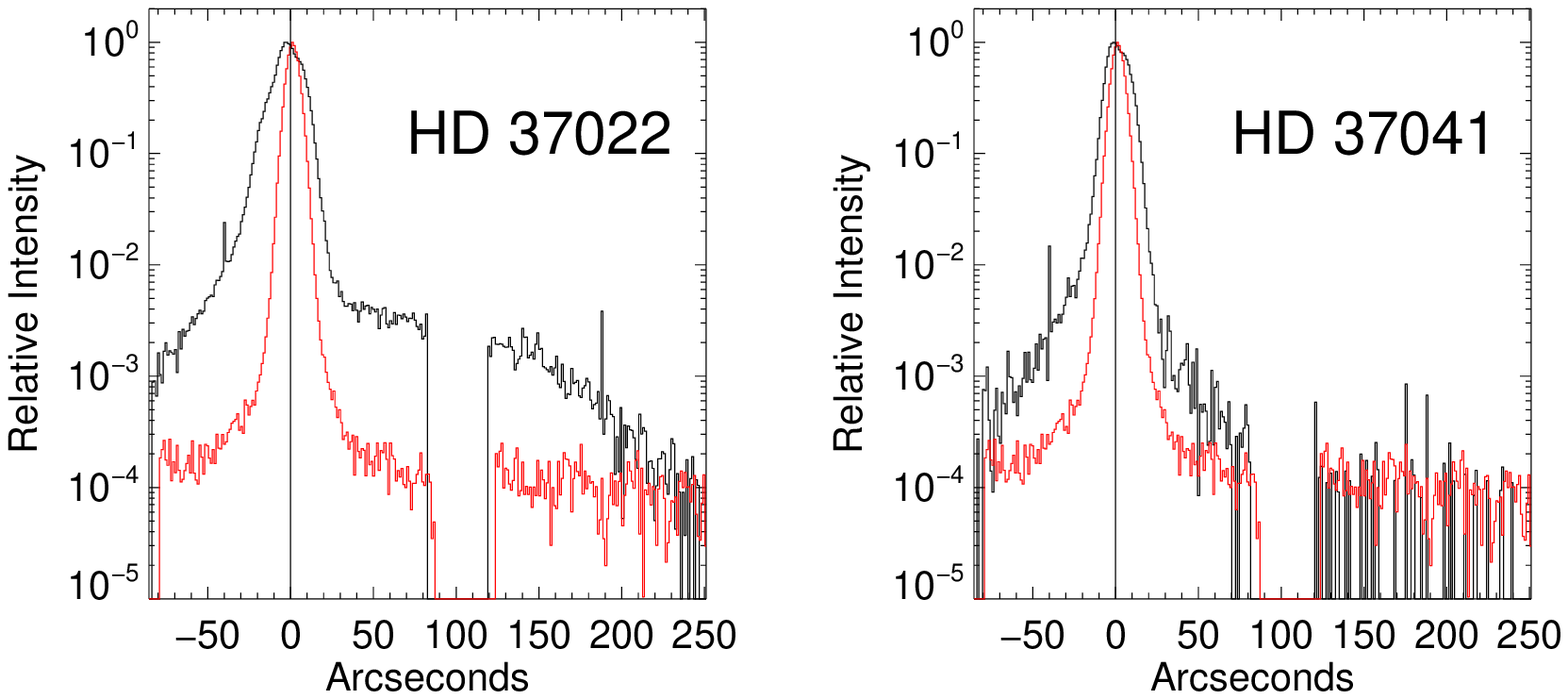}
   \includegraphics[width=5.5cm,clip]{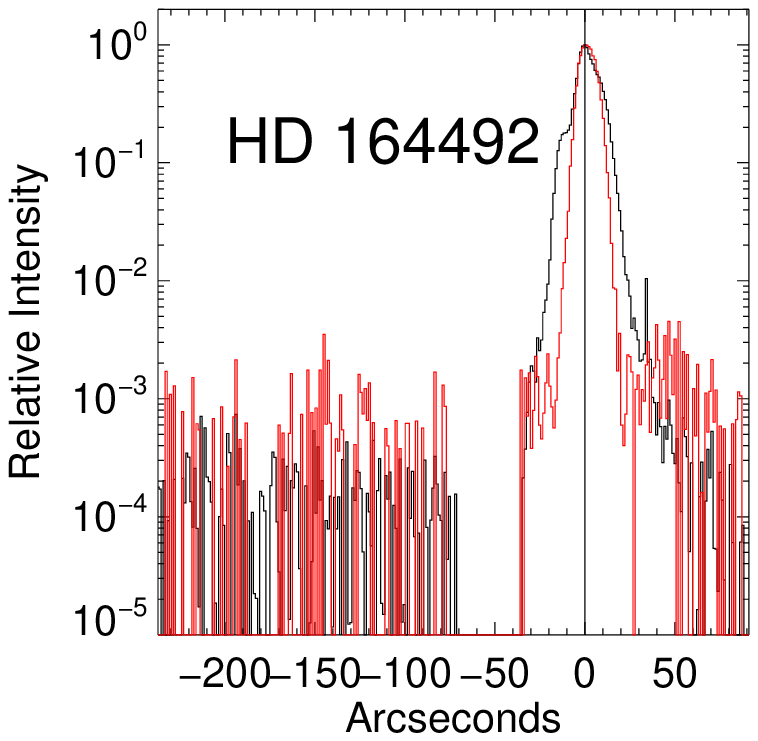}
   \end{tabular}
   \end{center}
   \caption[example] 
   { \label{scat} 
Observations of stellar scattered light profiles and nebular emission (black) compared to laboratory measurements of point source scattered light profiles (red).}
   \end{figure}

As mentioned in the previous section, the combined mirror reflectivity dropped by a factor of two between telescope build-up and post-flight measurements. Changes in the end-to-end throughput due to degradation have been monitored using the soft bremsstrahlung emission generated by an electron-impact soft X-ray lamp mounted on the telescope shutter door. During consecutive runs at JHU we found that this procedure is sensitive to the payload pressure at the time of the test, as well as to variations in filament current, both of which are not reproducible. Example lamp spectra are shown in Figure~\ref{xlamp}. Overall, this experiment shows no evidence for throughput degradation due to contamination during travel. We conclude that the decrease in mirror reflectivity occurs before the end-to-end testing, reaching a value that does not vary significantly afterwards. 

The combination of the Celestial ACS and the ST-5000 star-tracker allowed for a fast target acquisition, maximizing the observing time. We performed an in-depth analysis of the pointing accuracy and stability in order to constrain the sky coordinates associated with our observations during the 36.243UG flight. For this purpose we used the video recorded by the on-board Xybion camera. For each pointing step delimited by uplink maneuvers the Xybion frames within the corresponding time interval were co-added. Figure~\ref{xybion} shows the resulting image at the time of the observation of $\theta^1$~Ori~C overlaid on a H-band image of the same field from the 2MASS archive, showing the star positions and the misalignment between the experiment optical axis and the telescope. The Xybion composite has been cropped to emphasize the Trapezium region. Two stars in the Xybion field of view of 20\arcmin\ were usable for centroid calculations. The pointing of the telescope was determined by associating the absolute coordinates of the stars with their centroids. These values were then compared to the celestial coordinates provided by the star-tracker. The difference between them represents the misalignment between the star-tracker and the telescope. This is shown in Figure~\ref{acsoff} in terms of pitch and yaw offsets for all the time intervals between maneuvers. We note a good and relatively stable alignment along the pitch axis, but find evidence for a drift in alignment along the yaw axis. The gap in the data corresponds to a change of targets involving a 90\deg\ roll. Overall, the alignment didn't vary by more than 30\arcsec, within requirements of $\pm$1\arcmin. We also investigated the pointing stability between maneuvers. The Xybion frames were co-added in smaller subsets, varying from 1 to 50 images, to ensure that the results are not sensitive to bad centroids. In Figure~\ref{pshift} we show the comparison between the pointing stability derived from the measured positions of the two reference stars relative to the expected ones (upper plots), and the pointing stability recorded in the ACS housekeeping data (lower plots), for two representative time intervals between maneuvers. In both cases we identify attitude changes not detected or compensated by the ACS. The first interval shows a steady drift in the yaw axis, reaching 3\arcsec\ in 43~s, and the second interval includes a sudden jump by about 3\arcsec\ in pitch and 6\arcsec\ in yaw, while the reported ACS pointing errors remain smaller than  1\arcsec. This is a robust result, as the centroids of both stars (shown in black and red in the top row of Figure~\ref{pshift}) exhibit the same behavior. The jump in position at T+366~s was confirmed by a decrease in the count rate of the MCP and a shift in the image recorded by the CCD, and is correlated with a jump in the difference between the declination value reported by the star-tracker and that inferred from the Xybion images.

\section{Flight Data}

In this section we present preliminary extracted spectra from the second and third LIDOS flights. The results from the 36.208UG flight have been described in Ref.~\citenum{France:05}. The MCP data is divided into time intervals corresponding to stable count rates between maneuvers. The images were corrected for non-linear distortions and rotations, dark subtracted and flat fielded. The CCD exposures were corrected for slit rotation and the spectral extraction region was dark subtracted using a median of three non-illuminated areas from the same image. Large variations due to cosmic ray hits were removed by comparing the signal variation in adjacent pixels. Uncertainties related to vignetting features due to the repeller grid at the entrance of the spectrograph require a more detailed modeling, dependent on the spectral signal strength. During both flights the stellar field contained closely spaced stars, falling together within the slit aperture. A secondary peak is observed in the CCD spectra of $\theta^1$~Ori~C and in the MCP spectra of $\theta^1$~Ori~C, and HD~164492, the illuminating star of the Trifid nebula. In Figure~\ref{scat} we show the MCP spatial profiles for three pointings, containing $\theta^1$~Ori~C (HD~37022) and $\theta^2$~Ori (HD~37041) in Orion, and HD 164492 in the Trifid nebula, normalized to the maximum height. The profiles were obtained by collapsing the 2D MCP spectrum along the spatial direction after correcting for distortions. For comparison, laboratory scattered light profiles are over-plotted in red. The laboratory data was taken by offsetting the 20~$\mu$m pinhole a the entrance of the vacuum collimator in the vertical direction by an amount corresponding roughly to the position of the star in the slit during flight, and along the optical axis by the amount required to reproduce the telescope focus position during flight. Aside from the presence of a secondary source, revealed by the shoulder of the main feature in two of the plots, we note the presence of dust-scattered stellar light in the Orion observations, but also the lack thereof in the Trifid data. The wavelength dependence of the scattered light relative to the stellar flux will reveal information about the dust properties in these objects\cite{France:04,Burgh:02}. nebular spectra will be extracted from all pointings and the emission properties will be correlated with the distance from the illuminating stars.

\begin{figure}[t]
   \begin{center}
   
   \includegraphics[width=16cm,clip]{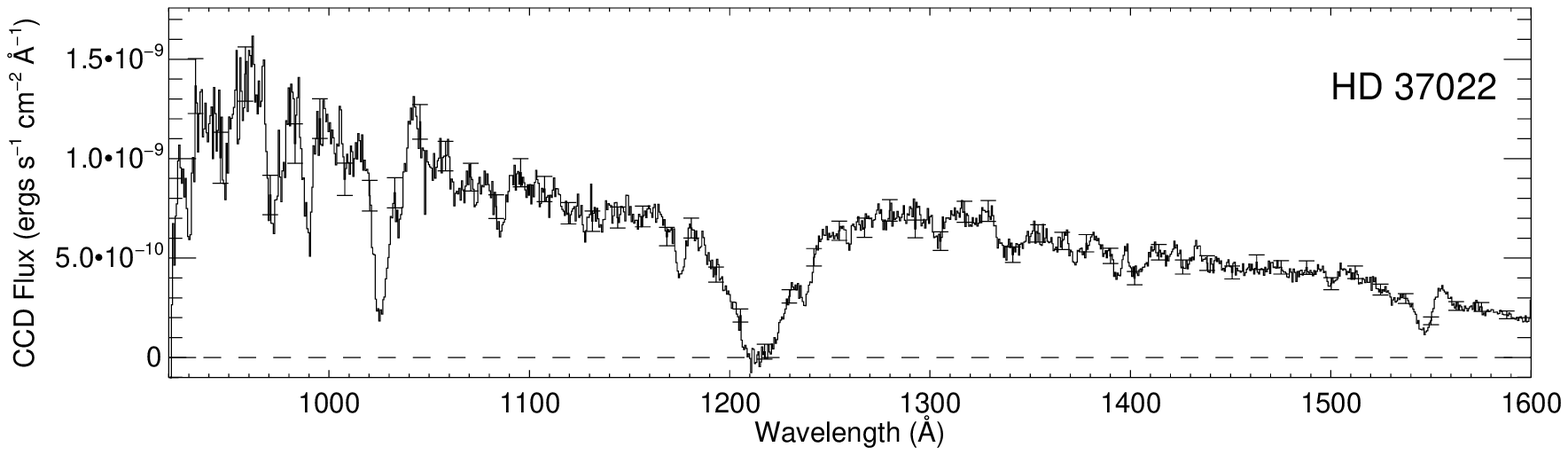}
    \includegraphics[width=16cm,clip]{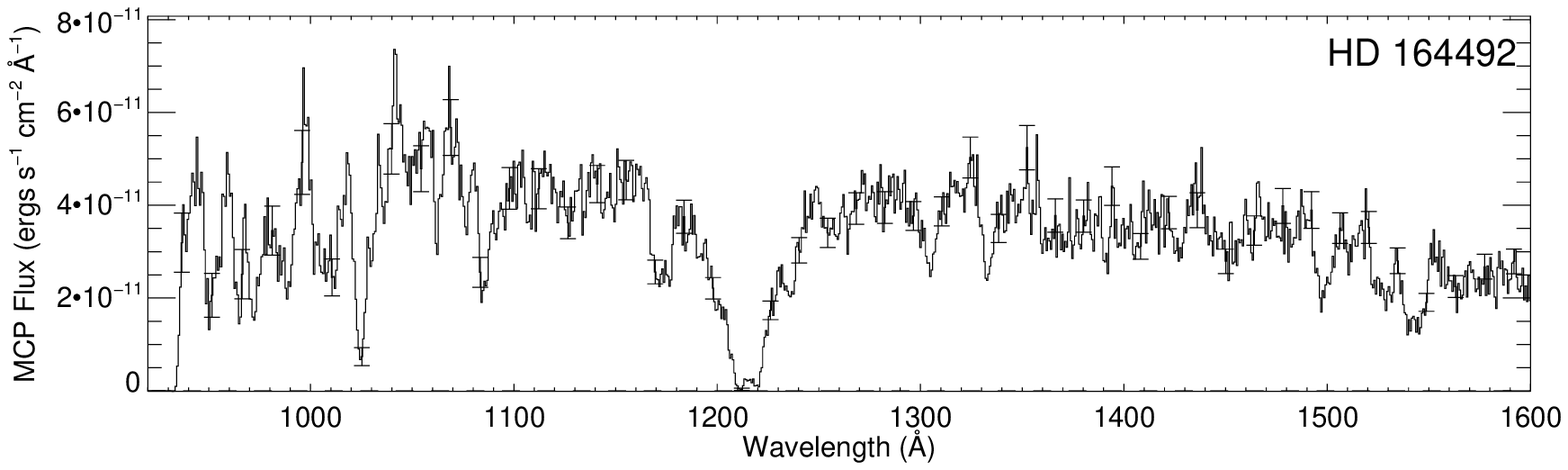}
   
   \end{center}
   \caption[example] 
   { \label{flx} 
Preliminary spectral extraction and calibration for $\theta^1$~Ori~C (upper panel), and HD 164492, the illuminating star of the Trifid nebula (lower panel).}
   \end{figure} 
  
The $\theta^1$~Ori acquisitions clearly show contributions from components C and A. We devised a custom procedure for spectral extraction, to avoid contamination from stellar companions. We separated the spectrum in half along the spectral direction and integrating over the uncontaminated half in the spatial direction. At longest wavelengths the two spectra were well separated, allowing for a good estimate of the correction needed by the extraction method to account for the light lost in the other spatial half of the spectrum. The spatial integration region was varied as a function of wavelength to account for the change in the height of the spectrum and avoid adding unnecessary background. All CCD exposures and MCP acquisition time intervals were processed separately, and then the resulting spectra were co-aligned and co-added. Figure~\ref{flx} shows the current estimates for the stellar fluxes for $\theta^1$~Ori~C recorded by the CCD during 36.243UG (top), and HD~164492 recorded by the MCP during 36.220UG (bottom). The flux calibration was performed using the effective area curves presented in \S~3. The absolute scale is subject to ongoing verification, by tracking the shadow cast on the CCD and constraining the MCP linearity curve. The last point is especially important for the Orion observations, when the count rate was as high as 29$\times$10$^{3}$~counts~s$^{-1}$. The effects of vignetting on the CCD spectra will be estimated using the spectral image for the isolated star HD~37041. The final objective is to provide reliable values for the spectral energy distribution of the observed young hot stars at far-UV wavelengths and use them to verify the predictions of theoretical photodissociation region models in conjunction with ancillary data at optical, infrared, and X-ray wavelengths.

\section{Conclusions}

The field of far-UV spectroscopy is continuously seeking innovative designs to maximize the throughput and dynamic range for astrophysical observations. As proven by the success of the $FUSE$ and the $Galaxy$ $Evolution$ $Explorer$, many research areas benefit form the understanding of molecular and atomic signatures in this bandpass, including Solar System objects, star formation, galaxy evolution, and cosmic reionization. A good handle on the processing of UV radiation in the local universe translates into better predictions and observing plans for objects at high redshift. Obtaining high quality data in this spectral range is limited not only by the necessity of using space-based instruments, as to avoid atmospheric absorption, but also by the low efficiencies of the detectors and coatings available. By incorporating new design concepts, the LIDOS instrument has given us the possibility to test and characterize recent developments in technology within the limited resources available for a rocket flight.  

The comprehensive analysis of the performance and characteristics of the LIDOS instrument has shown a good agreement with the design expectations. The total effective area and image quality was limited by the performance of the telescope. The pre-flight calibrations have been essential in determining the observing plan and setting the working parameters for the detectors, while customized post-flight studies provided better constraints for data reduction. The rocket experiments have proven that the dual-order design is a viable option for future far-UV instruments, and the combined use of the two detectors is efficient in providing redundancy and a significantly increased dynamic range. The CCD detector has an excellent uniform response across the chip, very good image fidelity, and a very high saturation limit. As such, a CCD spectrograph would benefit from the increase in throughput in a two-bounce design, or from a grating blaze transferring power from the symmetric order. On the down side, its ability to detect faint sources is limited by read noise and dark current. The CCD performance in uniformly illuminated fields is degraded somewhat by a red leak. Also, in our launch environment, the necessity to operate at a low temperature puts constraints on launch preparations and observing time. The MCP is a robust detector, with a very low background level, but becomes quickly non-linear and manifests poor imaging performance compared to the CCD. The pixel mapping suffers from distortions in both spatial and spectral directions, and the measured line spread functions show a scatter in pixel distribution much higher than expected. However, the photon counting ability of the MCP with its instantaneous response provides an unrivaled advantage in detecting the faintest emissions and making real-time observing decisions. An ideal detector would combine the imaging performance and linearity of the CCD with the low background and photon counting ability of the MCP. 

Overall, in spite of the identified shortcomings pertaining to cooling issues, focus consistency, and contamination control, the design of the mission and instrument have proven to be fault tolerant. An in-depth analysis of the instrument capabilities and associated error margins has led to the acquisition of valuable scientific information.
Preliminary analysis of the flight data has been performed, but the final data products are not yet obtained. We will seek further clarification on the absolute flux scale of the stellar spectra, and reliable estimates for the background and slit orientation for the extraction of diffuse nebular emission. We believe that our observations will contribute to constructing better models for the interaction between UV radiation and dust particles in the interstellar medium.

\acknowledgments
The author would like to thank Russell Pelton, the associated engineer, for his extensive support during all phases of payload build-up, calibration and testing, as well as the maintenance of an operational rocket laboratory. We also thank NSROC and the US Navy for the subsystems, facilities, and all the people who contributed to the successful launches of 36.208UG, 36.220UG, and 36.243UG. 
This work was supported by NASA grants NNG04WC03G and NNX08AM68G to the Johns Hopkins University.

\bibliography{mypaper}   
\bibliographystyle{spiebib}   

\end{document}